# Monte Carlo event generators for high energy particle physics event simulation

**Edited by:**
Andy Buckley[21], Frank Krauss[3], Simon Plätzer[27], Michael Seymour[23]
**on behalf of:**
Simone Alioli[16], Jeppe Andersen[3], Johannes Bellm[11], Jon Butterworth[18], Mrinal Dasgupta[23], Claude Duhr[1,18], Stefano Frixione[6], Stefan Gieseke[9], Keith Hamilton[18], Gavin Hesketh[18], Stefan Hoeche[14], Hannes Jung[2], Wolfgang Kilian[26], Leif Lönnblad[11], Fabio Maltoni[18], Michelangelo Mangano[1], Stephen Mrenna[4], Zoltán Nagy[2], Paolo Nason[16], Emily Nurse[18], Thorsten Ohl[28], Carlo Oleari[16], Andreas Papaefstathiou[13,19], Tilman Plehn[5], Stefan Prestel[11], Emanuele Ré[1,10], Juergen Reuter[2], Peter Richardson[1,3], Gavin Salam[25], Marek Schoenherr[3], Steffen Schumann[22], Frank Siegert[15], Andrzej Siódmok[7], Malin Sjödahl[11], Torbjörn Sjöstrand[11], Peter Skands[12], Davison Soper[24], Gregory Soyez[8], Bryan Webber[20]

[1] CERN, [2] DESY, Germany, [3] Durham University, UK,
[4] Fermi National Accelerator Laboratory, USA, [5] Heidelberg University, Germany,
[6] INFN, Sez. di Genova, Italy, [7] Institute of Nuclear Physics, Cracow, Poland,
[8] IPhT Saclay, France, [9] Karlsruhe Institute of Technology, Germany,
[10] LAPTh Annecy, France, [11] Lund University, Sweden, [12] Monash University, Australia,
[13] NIKHEF, Netherlands, [14] SLAC National Accelerator Laboratory, USA,
[15] Technical University of Dresden, Germany, [16] Università di Milano-Bicocca, Italy,
[17] Université Catholique de Louvain, Belgium, [18] University College, London, UK,
[19] University of Amsterdam, Netherlands, [20] University of Cambridge, UK,
[21] University of Glasgow, UK, [22] University of Göttingen, Germany,
[23] University of Manchester, UK, [24] University of Oregon, USA,
[25] University of Oxford, UK, [26] University of Siegen, Germany,
[27] University of Vienna, Austria, [28] Würzburg University, Germany

**Abstract**

Monte Carlo event generators (MCEGs) are the indispensable workhorses of particle physics, bridging the gap between theoretical ideas and first-principles calculations on the one hand, and the complex detector signatures and data of the experimental community on the other hand. All collider physics experiments are dependent on simulated events by MCEG codes such as Herwig, Pythia, Sherpa, POWHEG, and MG5_aMC@NLO to design and tune their detectors and analysis strategies. The development of MCEGs is overwhelmingly driven by a vibrant community of academics at European Universities, who also train the next generations of particle phenomenologists. The new challenges posed by possible future collider-based experiments and the fact that the first analyses at Run II of the LHC are now frequently limited by theory uncertainties urge the community to invest into further theoretical and technical improvements of these essential tools. In this short contribution to the European Strategy Update, we briefly review the state of the art, and the further developments that will be needed to meet the challenges of the next generation.

# Monte Carlo Community input to European Strategy Update

**Executive Summary**

Monte Carlo event generators (MCEGs) are the indispensable workhorses of particle physics, bridging the gap between theoretical ideas and first-principles calculations on the one hand, and the complex detector signatures and data of the experimental community on the other hand[1]. All collider physics experiments are dependent on simulated events by MCEG codes such as Herwig[2], Pythia[3], Sherpa[4], POWHEG[5], and MG5_aMC@NLO[6] to design and tune their detectors and analysis strategies. The development of MCEGs is overwhelmingly driven by a vibrant community of academics at European Universities, who also train the next generations of particle phenomenologists.

The new challenges posed by possible future collider-based experiments and the fact that the first analyses at Run II of the LHC are now frequently limited by theory uncertainties urge the community to invest into further theoretical and technical improvements of these essential tools:
- systematic inclusion of emerging fixed-order calculations at next-to-next-to-leading order (NNLO) in QCD and mixed QCD-electroweak calculations at second order in perturbation theory or higher; in addition, in preparation for a possible electron-positron collider, inclusion of calculations accurate to NNLO or better in the electroweak theory;
- systematic improvement of parton showers to achieve a parametric accuracy that is equivalent to next-to-next-to-leading logarithmic (NNLL) order in resummation for a wide range of observables, the inclusion of terms corresponding to sub-leading colour corrections, and the combination of fixed-order calculations with the parton shower at increasing precision;
- connecting parton shower methodology and analytic resummation and exploring new structures for parton showers such as amplitude-level evolution;
- augmenting phenomenological models for non-perturbative effects such as the transition of partons to hadrons and multiple partonic interactions, and underpinning them with better theoretical foundations;
- bridging the persistent gaps to other communities using event generators in different physical settings: non-collider based experiments and heavy ion physics;
- adapting the software frameworks to exploit modern high-performance computing hardware environments and increase the efficiency of CPU intensive high-precision event generation;
- maintaining and expanding frameworks for rapid verification of the programs, validating their physical agreement with data, and optimizing the parameters of their phenomenological models.

**Event Generator Physics Community**

The development of the major multi-purpose event generator packages is almost entirely located in European universities, as well as CERN, and co-ordinated through a Horizon 2020 Initial Training Network, MCnetITN3. Other important programs for individual parts of the event generator chain are also developed outside this network, again largely in European universities. Close links also exist to institutions in Australia and the U.S., in many cases through the movement of European-trained researchers. The MCnet network provides a common backbone of exchange for many of these activities, and plays a significant role in the training of young researchers in both theoretical and experimental physics, through a well established series of international schools, and an innovative short-term studentship exchange scheme.

**Perturbative Improvements: Matching and Merging**

For the full exploitation of experiments it is mandatory that the results of ongoing and future work on calculations at highest precision, either at fixed order or with resummation techniques, are systematically embedded into the event generators. This requires that these techniques be combined with the resummation available in the parton showers in a theoretically consistent fashion (known as *parton shower matching*) as well as the combination of many matched samples with increasing final-state multiplicity into a single inclusive sample, a process known as *multijet merging*. In particular:

- High-Luminosity Run of the LHC and future hadron colliders:

  - The ongoing "NNLO revolution" has resulted in differential and total cross-sections for practically all relevant $2 \rightarrow 2$ scattering processes, which have not yet been systematically included into our codes. First steps, focusing on the production of colour-singlets, led to two matching algorithms which will be extended to also include colourful final states. We also expect results for more complicated scattering topologies, i.e. $2 \rightarrow 3, 4, ...$ processes to become available, and our matching algorithms will be capable to produce theoretically robust prediction.

  - In a subsequent step the NNLO-matched samples for towers of increasing jet multiplicities will be combined into inclusive samples, for example for the associated production of vector or Higgs bosons or boson pairs or of top quarks with jets. Drawing from our long experiences we will develop, validate and compare multijet merging algorithms.

  - To guarantee precision at the 10% level or better, also in the high-scale tails of distributions that are phenomenologically relevant for searches for new physics, electroweak corrections will be systematically included into our simulations to all relevant processes. Due to their calculational cost we will also need to provide process-independent fast estimates that are correct at the level of a few percent, as provided by the Sudakov approximation. For a number of standard candle processes, even mixed electroweak-strong corrections will become mandatory to ensure that important experimental data analyses are not limited by MC uncertainties.

- Future electron-positron colliders:

    - For future electron-positron colliders the analysis of data and interpretation of results will necessitate theoretical control at the sub-percent level for some processes, which can only be achieved by including multi-loop QED and the full 2-loop or better electroweak corrections. In turn, these will need to be included into the simulation when the lepton collider setups are revisited.

    - In addition, for the case of a linear collider setup, beamstrahlung must be added into the simulation.

    - Finally, some work on collisions of photons or of photons with leptons will be essential.

**Parton Shower Algorithms and Resummation**

- *Improved theoretical accuracy of parton showers: perturbative orders*
  Drawing on our experience with the matching and merging of NLO calculations with the parton shower, we anticipate that a theoretically robust procedure at NNLO will demand an improvement of the all-order perturbative precision of the shower. This will require inclusion of all terms of second order in the strong coupling currently elided from parton shower implementations. This signifies another step change towards precision simulations, and needs a step-change in the theoretical foundations of parton showers, which have, apart from very recent suggestions, not been significantly changed in the past 3 decades. The inclusion of second-order corrections will challenge the current probabilistic parton shower paradigm.

- *Improved theoretical accuracy of parton showers: sub-leading colours*
  Another important improvement in the theoretical accuracy of parton showers will come from a careful analysis of the underlying paradigm of keeping leading colour contributions only. Parametrically we expect sub-leading contributions to contribute at the level of $1/N_c^2$, i.e. around 10%. There are prospects for significant changes in important signatures such as rapidity gaps (e.g. for EW vector boson fusion and scattering processes) when sub-leading colour corrections are systematically included to all orders and not just to fixed order. Including these effects also necessitates a significant change in the established probabilistic paradigm.

- *Electroweak parton showering*
  For hadron colliders at the highest energies, the "soft" emission of vector bosons will become a more dominant effect than at current colliders, where the available energies do not introduce a sufficiently large hierarchy of scales to mandate a meaningful resummation of leading logarithms to all orders. This will certainly change at a possible future 100 TeV machine, and it will mandate the inclusion of the chirality-dependent multiple emission of W and Z bosons off quarks — an effect that is currently mostly neglected. As a consequence, current parton showers will need to be re-engineered such that chirality information of the quarks can be transported without being computationally prohibitive.

- *Link to resummation*
  In the context of increasing the perturbative accuracy of parton showers, we will also investigate the link to analytic resummation techniques, to arrive at a better understanding of similarities and differences and their impact on a range of observables. The structure of analytic resummation will also be used in addressing the development of higher accuracy parton shower algorithms, and parton shower methods are expected to play an increasingly important role in dedicated resummation programs.

**Phenomenological Models**

- *Hadronization models*
  Current phenomenological hadronization models are based on assumptions and qualitative insights dating back to the 1970's, and advances have mainly been driven by comparison with data of increasing quality, especially from LEP, rather than new theoretical insight. The development of a more consistent validation and tuning strategy will include information from other data, for example DIS data taken at HERA, to get a better handle on effects such as the modification of fragmentation by (semi-)coloured initial states. This may result in further developments, e.g. in the handling of hadronization in hadron-hadron collisions.

- *Underlying Event*
  - The same necessity for theoretical improvements applies also to the description of multiple partonic interactions (MPI), which contribute significantly to the overall particle yield and to jet observables relevant in searches for new physics through boosted objects. Work on a better understanding of MPI based on first-principles QCD must be supplemented with better models which are also capable of capturing effects such as diffraction in one coherent framework. This will also feed back into hadronization models, linking them to the further development of models of inclusive particle production (i.e. "minimum bias" events), possibly motivated by perturbative QCD.

  - For experiments at electron-positron colliders it may also become necessary to include effects such as additional "multiple parton" interactions driven by photons from initial state radiation and/or beamstrahlung. The continued provision of reference data and parameter optimisation tools for the phenomenological models and model validation constitutes an important part of the overall simulation infrastructure.

- *Inclusive QCD and the Heavy Ion connection*
  Several observations by LHC, indicating similarities between high multiplicity proton-proton collisions and proton-ion/ion-ion ones, have challenged current models of soft QCD interactions, as well as phenomenological hadronization models. Thus, while current models must continue to be scrutinised and improved in order to provide better descriptions of the overall particle yields and jet observables — driven by the requirements of new-physics searches — the investigation of similarities between pp and AA physics is deeply interesting in its own right as a fundamental exploration of QCD.
  Monte Carlo event generators have traditionally played a smaller role in the pA/AA community than in the pp one, but a new avenue for event generators is opened up by the potential for an improved, microscopic understanding of the dynamics of the quark-gluon plasma (QGP) produced in AA collisions. For present investigations of ultra-peripheral nuclear collisions, and for future electron-ion and electron-positron experiments, MPI induced by photons must also be understood.

Diffractive and forward physics is another area in need of deeper understanding, with links to cosmic-ray physics. In summary, systematic improvement based both on models and first principles QCD is key to obtaining a unified picture of particle production and parton dynamics for all collision systems.

- *Event generators for low-energy physics (neutrinos and "beyond colliders" physics)*
  Similar to the heavy-ion community, non-collider experiments such as the neutrino experiments or dark matter searches often face challenges that are absent in collider physics, and thus commonly rely on specific event generator tools that are sometimes unique to individual experiments. This fragmentation of the community will be overcome, with intensified cross-talk, for example in hadronization models or the simulation of hadron decays, between event generators deployed in low- and high-energy experiments.

All these developments in the phenomenological modelling components of MCEGs are reliant — to a greater extent than for the first-principles perturbative physics — on the availability of mechanisms for validating the models and tuning their parameters. Delivery of high-precision MC predictions requires that analyses from previous experiments be made available through public analysis platforms such as Rivet and MadAnalysis to enable quick validation and tuning.

**Development of Methodology**

Apart from direct physics improvements, it is also necessary to overhaul the overall structure of the leading MCEG programs to maximise the effectiveness of their deployment on modern multi-core parallel or vector computing architectures. In parallel, the various stages of event simulation must become more efficient to enable large-scale production of statistically relevant samples. This is particularly critical for future collider programmes as all the above areas with potential for improvement in the field-theoretical accuracy of MCEG modelling will come with a high cost in algorithmic complexity — perhaps prohibitively so without significant investment in software engineering and high-performance algorithm development. New developments starting "from scratch" will directly address future computing models as far as they can be anticipated today.

A significant success of recent years has been the development of a variety of shared "community" components of the MCEG ecosystem, for example utility libraries such as LHAPDF and HepMC, and validation and tuning frameworks such as Rivet[7] and Professor. These bring efficiencies in the use of available effort, since they avoid duplication of component development. However, this brings new challenges in their support, for which mechanisms need to be developed to share the effort between event generator projects and their user communities. The imperative to ensure computational efficiency extends to these packages, some of which currently have significant performance bottlenecks and, again, investment of effort by the development and user communities as whole is needed.

## Challenges in the development of the next generation

Currently the community of scientists developing event generators at high energies is in a healthy state, thanks to active research groups, mainly at European Universities, and relatively consistent funding in the past decade, provided through a sequence of successful ITN's and ETN's. This funding has also facilitated a string of successful annual international schools, which started in 2006 and have been running ever since.

To address the challenges in developing high-precision perturbative components and new phenomenological models, and ensuring their deployment in modern computing environments, it is essential that this funding be maintained and enhanced. The particular profile of the community that develops MCEGs needs to be recognized, as a core activity in theoretical physics, but which contributes to planning the next generation of experiments and as an essential facilitator to experimental analyses. Ground-breaking advances, including the agenda summarized in this document, will only be possible if the development and exploitation of event generators is fully recognized as an independent research discipline within particle physics and quantum field theory, with its research being vital to moving the field forward even in absence of concrete experimental demands.

## Acknowledgement

The MCnet project has received funding from the European Union's Horizon 2020 research and innovation programme under the Marie Skłodowska-Curie grant agreement No 722104 "MCnetITN3".